\documentclass[
superscriptaddress,
nofootinbib,
twocolumn,
 amsmath,amssymb,
 aps,
pra,
notitlepage,
longbibliography
]{revtex4-1}

\usepackage{dcolumn}
\usepackage{bm}
\usepackage{epsfig}
\usepackage{graphicx}
\usepackage{latexsym}
\usepackage{amsfonts}
\usepackage{setspace}
\usepackage{graphicx}
\usepackage{amsmath}
\usepackage{mathrsfs}
\usepackage{verbatim}
\usepackage{color}
\usepackage{SIunits}
\usepackage{hyperref}

\usepackage{physics}

\newcommand{\be}{\begin{equation}}
\newcommand{\ee}{\end{equation}}
\newcommand{\ba}{\begin{array}}
\newcommand{\ea}{\end{array}}
\newcommand{\bqa}{\begin{eqnarray}}
\newcommand{\eqa}{\end{eqnarray}}

\begin{document}

\title{A rigid, low-loss fiber-optic coupler for cryogenic photonics}

\author{Mengdi Zhao} 
\affiliation{Holonyak Micro and Nanotechnology Laboratory, University of Illinois at Urbana-Champaign, Urbana, IL 61801 USA}
\affiliation{Illinois Quantum Information Science and Technology Center, University of Illinois at Urbana-Champaign, Urbana, IL 61801 USA}
\affiliation{Department of Physics, University of Illinois at Urbana-Champaign, Urbana, IL 61801 USA}
\author{Kejie Fang} 
\email{kfang3@illinois.edu}
\affiliation{Holonyak Micro and Nanotechnology Laboratory, University of Illinois at Urbana-Champaign, Urbana, IL 61801 USA}
\affiliation{Illinois Quantum Information Science and Technology Center, University of Illinois at Urbana-Champaign, Urbana, IL 61801 USA}
\affiliation{Department of Electrical and Computer Engineering, University of Illinois at Urbana-Champaign, Urbana, IL 61801 USA}

\begin{abstract} 

Recent developments in quantum light-matter coupled systems and quantum transducers have highlighted the need for cryogenic optical measurements. In this study, we present a mechanically-rigid fiber-optic coupler with a coupling efficiency of over 50\% for telecom wavelength light at cryogenic temperatures. Our method enables sensitive photonic device measurements that are alignment-free and immune to mechanical vibrations in cryogenic setups.

\end{abstract}

\maketitle

Optical measurements are essential for a number of emerging quantum platforms and technologies operating in cryogenic environments, such as quantum optomechanics \cite{riedinger2016non, maccabe2020nano}, quantum microwave-to-optical transducers \cite{fan2018superconducting, mckenna2020cryogenic, mirhosseini2020superconducting, hease2020bidirectional}, and cryogenic electro-optic modulators \cite{eltes2020integrated,youssefi2021cryogenic}. These quantum systems coupled to optical fields consist of low-frequency mechanical oscillators and microwave resonators typically in the MHz to GHz frequency range. To preserve the quantum coherence of the mechanical and microwave resonators, it is necessary to operate these systems at cryogenic temperatures to suppress the thermal excitations. Even for certain photonic device measurements, cryogenic temperature are necessary to minimize parasitic thermomechanical and thermorefractive noises \cite{zhao2022observation}. Therefore, a low-loss fiber-optic coupler compatible with cryogenic environments is critical for the optical interrogation of these chipscale quantum systems.

Several cryogenic fiber-optic coupling methods have been developed, such as lens-fiber edge coupling \cite{cohen2015phonon}. However, this method requires a specific device fabrication process and a challenging blind alignment of the fiber and device inside a dilution refrigerator. Even if an optical window is available, fiber-device alignment can be difficult due to the resolution-limited imaging system placed outside the cryostat. Additionally, cryogenic optical measurements face challenges caused by mechanical vibrations generated by the pulse-tube cryostat. These vibrations can cause fluctuations in the fiber-optic coupling efficiency, which can affect sensitive optical measurements. To overcome these challenges, an angle fiber coupled with a grating coupler using optical adhesives has been demonstrated with an efficiency about 25\% for telecom light \cite{mckenna2019cryogenic}. However, this structure requires a long tapered region on the chip to interface the grating coupler and single-mode waveguide due to the substantial size mismatch of the two components. Grating couplers also have a limited bandwidth. Here, we demonstrate a mechanically rigid fiber-optic coupler with an efficiency of over 50\% in the telecom wavelength band, based on a specially-glued, adiabatically-tapered fiber coupler. This broadband coupler addresses the limitations of existing methods for cryogenic fiber-optic coupling, ensuring mechanical stability while maintaining a compact device footprint.

We demonstrate our method using the InGaP integrated photonic platform. The InGaP photonic devices are fabricated from a disordered In$_{0.48}$Ga$_{0.52}$P thin film grown on a GaAs substrate using metal-organic chemical vapor deposition. Due to the similar refractive index of InGaP and GaAs, we have developed an ``insulator-on-top" approach to create suspended InGaP photonic circuits \cite{zhao2022ingap}, which avoids the need to transfer the InGaP thin film to a low index substrate \cite{martin2017gainp}. To accomplish this, we deposit a thin layer of Al$_2$O$_3$ or SiO$_2$ on the patterned InGaP-on-GaAs chip via atomic layer deposition. The InGaP photonic circuit is then released from the GaAs substrate through selective wet etching of GaAs and is mechanically anchored to the insulator membrane. The detailed fabrication process is described in \cite{zhao2022ingap}. The insulator membrane also provides a platform to facilitate the positioning of the tapered fiber and gluing it to the waveguide coupler, resulting in a mechanically rigid fiber-optic coupler. Our ``insulator-on-top" method is applicable for creating integrated photonic circuits in general index-matched slab-on-substrate material systems without the need for the more complicated thin-film transfer process.

\begin{figure}[!htb]
\begin{center}
\includegraphics[width=1\columnwidth]{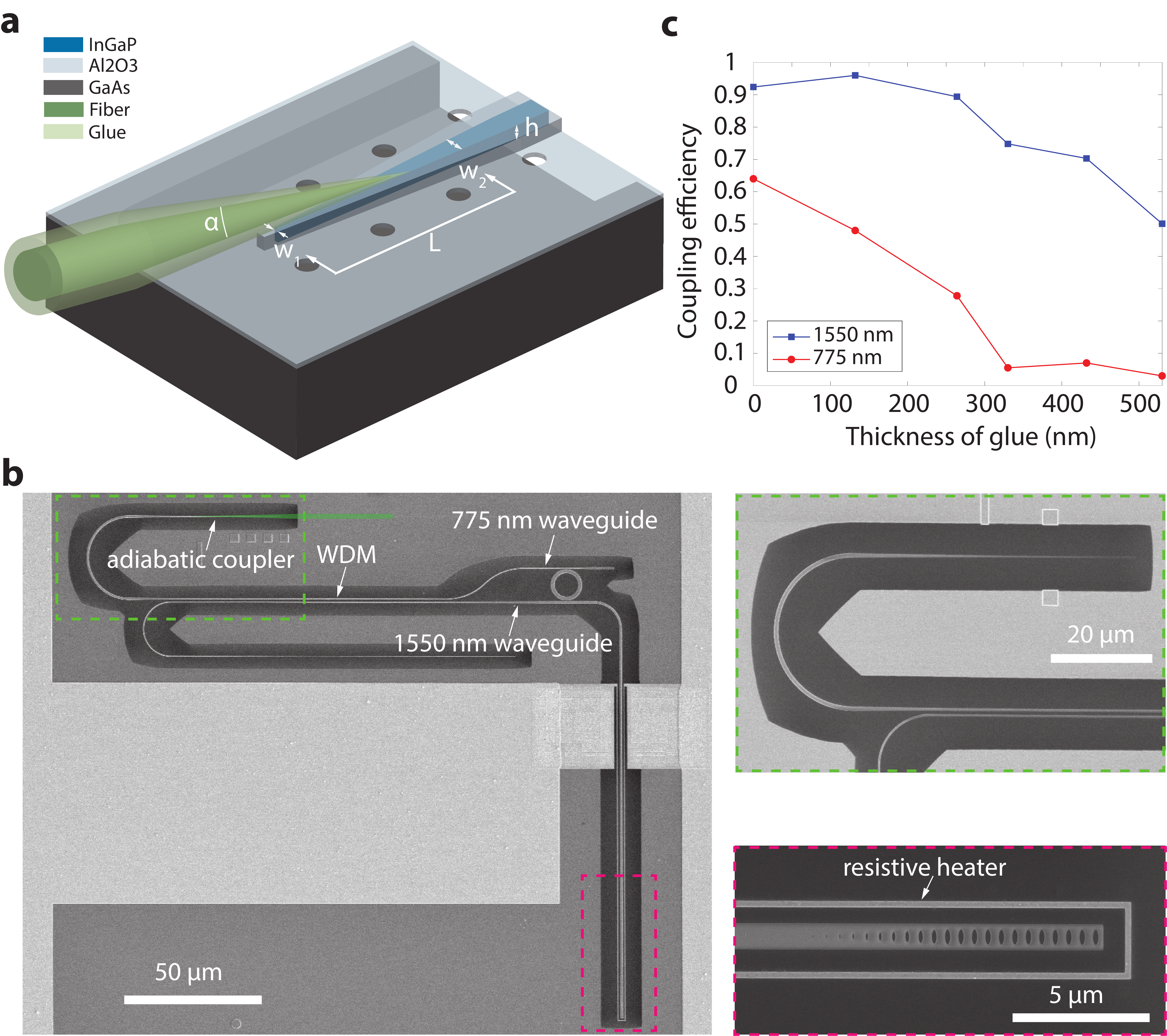}
\caption{\textbf{a}. Illustration of the adiabatic fiber-optic coupler mechanically fixed using optical adhesives. \textbf{b}. Scanning electron microscope images of the InGaP integrated photonic circuit. \textbf{c}. Simulated fiber coupler efficiency for 1550 nm TE and 775 nm TM light with adhesive of different thickness. }
\label{fig:device}
\end{center}
\end{figure}

Fig. \ref{fig:device}a and b show the schematic of the adiabatic fiber-optic coupler and scanning electron microscope images of the photonic circuit, respectively. The InGaP layer is 115 nm thick, and the Al$_2$O$_3$ membrane is 35 nm thick. The adiabatic coupler is designed to function for both the 1550 nm band transverse electric (TE) mode and 775 nm band transverse magnetic (TM) mode, which are relevant for second-order nonlinear optical effects in InGaP photonic circuits. The waveguide coupler is tapered from 50 nm to 320 nm width over 24 $\mu$m length. The tapered optical fiber with a 4$^{\circ}$ aperture is fabricated by hydrofluoric acid etching. Fig. \ref{fig:device}c shows the simulated coupling efficiency for the adiabatic coupler coated with a thin layer of optical adhesive ($n=1.56$). The coupling efficiency of the 1550 nm light drops slightly in the presence of the glue while the 775 nm light is subject to more losses because of the smaller mode volume. As shown in Fig. \ref{fig:device}b, the complete photonic circuit contains a 1550 nm/775 nm wavelength division-multiplexer (WDM) and separate 1550 nm and 775 nm band waveguides coupled with a microring resonator. The waveguides are terminated with photonic crystal mirrors. A resistive heater is fabricated adjacent to the 1550 nm band waveguide to control the transmission of the 1550 nm band light resonant with the microring \cite{zhao2023photon}.

\begin{figure}[!htb]
\begin{center}
\includegraphics[width=1\columnwidth]{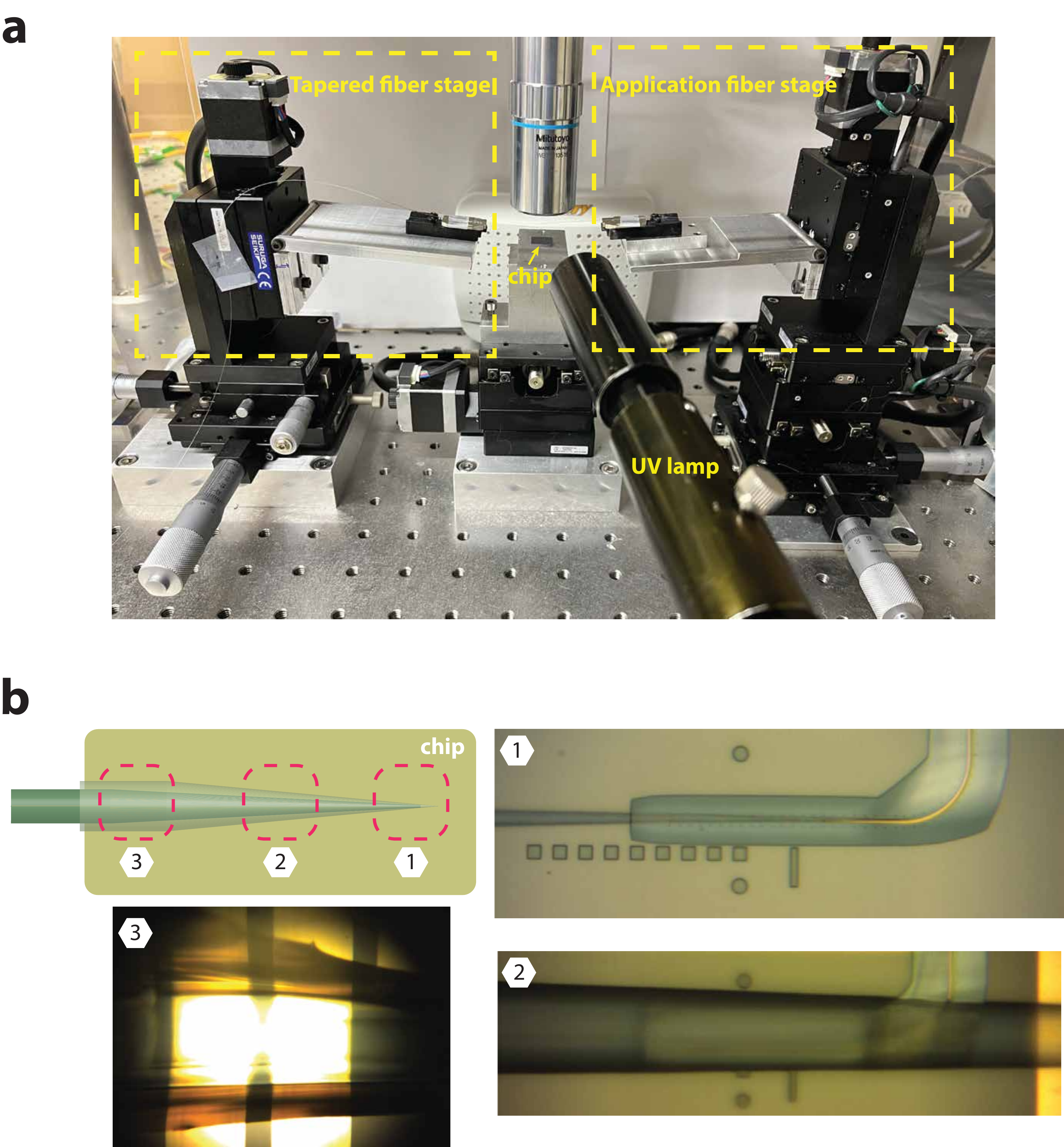}
\caption{\textbf{a}. The setup used for gluing the tapered fiber to the on-chip waveguide device. \textbf{b}. Optical microscope images of the glued fiber.  }
\label{fig:setup}
\end{center}
\end{figure}

The setup used for gluing the tapered fiber-optic coupler is shown in Fig. \ref{fig:setup}a. To prevent any movement during the gluing process, the chip is taped onto the sample stage. The tapered fiber is mounted on a fiber holder that is fixed on a set of XYZ-stages. The mounted fiber has a tilt angle of approximately 5 degrees with respect to the horizontal direction. Two types of UV optical adhesives, NOA 61 ($n=1.56$) and NOA 88 ($n=1.52$), are used for the gluing process. NOA 61 is utilized to glue the tapered fiber tip due to its oxygen inhibition-free properties, which enables a thin layer of adhesive at the fiber tip to be fully cured under UV exposure. On the other hand, large drops of NOA 88 adhesive are used to secure the bulk optical fiber to the chip because of its low outgassing under high vacuum conditions.

First, we glue the tapered fiber tip to the waveguide. We use a bare fiber mounted on another set of motorized stage to pick up a drop of NOA 61 adhesive. The tapered fiber is moved into the drop of glue manually so that a length of approximately 100 $\mu$m of the tapered fiber is wetted with glue. The fiber tip is then aligned and placed on the waveguide coupler using the motorized stages, and the fiber is further lowered so that the glue at the tip extends along the fiber by a distance of about 500 $\mu$m. The reflection signal in the 1550 nm band from the photonic circuit is monitored during this process to maximize the coupling efficiency by slightly adjusting the fiber position. The optimal coupling efficiency with the attached glue is about 50\%. The glue is cured under a UV light spot lamp (Dymax BlueWave 200) with an output power of 17 W/cm$^2$ in the UVA range for 3 minutes.

After the fiber tip is fixed to the device, a drop of NOA 88 is applied using the application fiber on the tapered fiber near the edge of the chip, which is about 2-3 mm from the fiber tip. 
The glue then flows along the gap between the fiber and the chip towards the tip of the fiber. The tilt angle of the tapered fiber is crucial for controlling the flow velocity. The flow of glue is monitored under the microscope together with the real-time reflection signal, so that the UV light is turned on promptly to stop the flow of the glue before it reaches a distance that the coupling efficiency starts to drop. The glue is then cured under the UV light for 3 minutes. Otherwise, if the drop of glue is cured immediately after application without extension of the glue, the thermal contraction of the cured glue bead during cool down will pull the fiber tip and detach it from the membrane. Finally, the glue is aged to achieve optimal adhesion by placing the chip on a 50 C heater for 12 hours in an ambient environment.

\begin{figure}[!htb]
\begin{center}
\includegraphics[width=0.7\columnwidth]{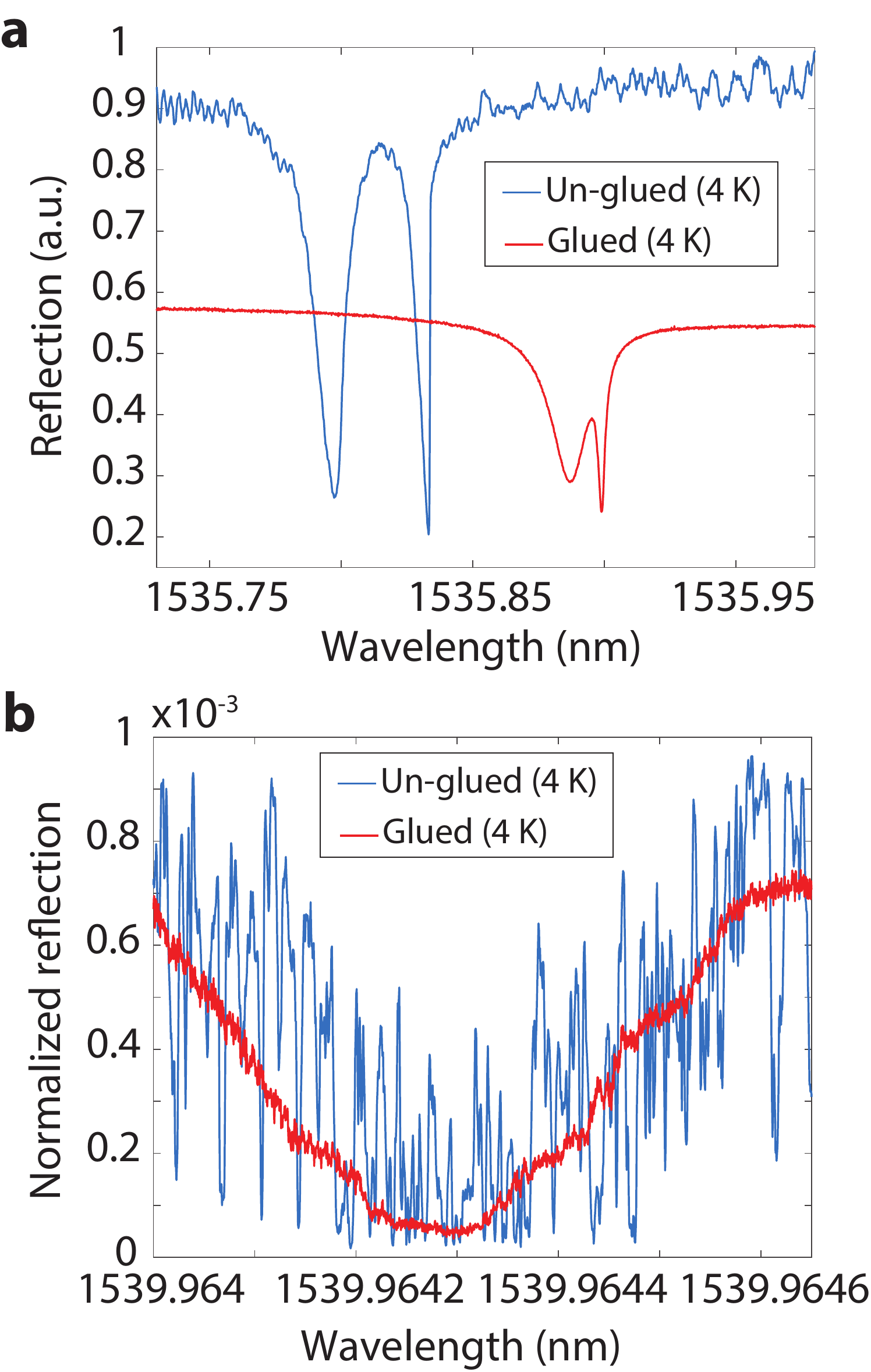}
\caption{\textbf{a}. Measured reflection spectrum of split resonances with and without glue at 4 K.   \textbf{b}. Measured reflection (normalized to the off-resonance reflection) near critical coupling with and without glue at 4 K. }
\label{fig:data}
\end{center}
\end{figure}

We measured the coupling efficiency of the adiabatic coupler at room temperature and 4 K. Prior to gluing, the coupling efficiency for the 1550 nm band TE light and 775 nm band TM light is approximately 70\% and 30\%, respectively. After gluing, the efficiency drops to 50\% and 2\% for the 1550 nm band and 775 nm band light, respectively. The coupling efficiency at 4 K remains the same as at room temperature, indicating the thermomechanical stability of the glued fiber-optic coupler. To demonstrate its mechanical rigidity, Fig. \ref{fig:data}a shows the reflection spectrum of microring resonators measured using free and glued tapered fibers at 4 K. The measurement using a free tapered fiber is facilitated by a piezo positioner in the dilution refrigerator with an optical window for fiber-device alignment. The vibration-induced coupling fluctuations seen in the spectrum measured using a free fiber are absent in the one measured using a glued fiber. The split-resonance can be tuned to the critical coupling condition by controlling the waveguide phase between the microring and the photonic crystal mirror using the resistive heater (Fig. \ref{fig:device}b) \cite{zhao2023photon}. Fig. \ref{fig:data}b shows the minimum reflection of a microring resonance near the critical coupling. The signal measured using the free fiber exhibits substantial fluctuations in contrast to the one measured with a glued fiber. The more severe fluctuations, in terms of the signal-to-noise ratio, in the reflected signal near critical coupling compared to the off-resonance reflection in Fig. \ref{fig:data}a is due to the fact that direct reflection at the tapered fiber coupler, which is about -35 dB, appears to be more sensitive to the fiber motion. We also measured the glued fiber-optic coupler at 20 mK and its performance is similar as 4 K. The yield of working glued fibers after cooling down to 4 K is larger than 90\%. However, we find that the bonding of the cured glue can only withstand a few temperature cycles.

To summarize, we have developed a mechanically-rigid fiber-optic coupler for high-efficiency coupling of light with devices at telecom wavelengths. The coupler was demonstrated to be thermomechanically stable and immune to mechanical vibrations in a cryogenic environment. The developed fiber coupling method is expected to have applications in sensitive optical measurements in cryogenic setups.

\vspace{2mm}
\noindent\textbf{Acknowledgements}\\ 
This work was supported in part by US National Science Foundation (ECCS-2223192) and U.S. Department of Energy Office of Science National Quantum Information Science Research Centers.

\end{document}